\input{aipcheck}

\documentclass[
    ,final            
  ]
  {aipproc}

\layoutstyle{6x9}

\usepackage{amsmath}
\usepackage{graphicx}
\usepackage{subfigure}

\begin{document}

\title{Finite-size effects on the hadron-quark\\ mixed phase}


\author{Tomoki~Endo}{
  address={Department of Physics, Kyoto University, Kyoto 606-8502, Japan}
}

\author{Toshiki~Maruyama}{
  address={Japan Atomic Energy Agency, Tokai, Ibaraki 319-1195, Japan 
}
}

\author{Satoshi~Chiba}{
  address={Japan Atomic Energy Agency, Tokai, Ibaraki 319-1195, Japan }
}

\author{Toshitaka~Tatsumi}{
  address={Department of Physics, Kyoto University, Kyoto 606-8502, Japan}
}

\begin{abstract}
We show that the hadron-quark mixed phase is restricted to narrow range 
 of baryon chemical potential
 by the charge screening effect. 
Accordingly the mixed phase expected in hadron-quark hybrid stars should
 be narrow.
Although the screening would not have
 large effect in bulk properties of the star such as mass or radius,
 it change the internal structure of the star very much, which may be
 tested by the cooling curve, glitch phenomena or gravitational waves.
\end{abstract}

\maketitle


\section{Introduction}

Recently it seems to be widely believed that the {\it structured mixed
phase} (SMP)
would appear in the wide density range 
during the first order phase transitions with many ($\geq2$) chemical
potentials \cite{gle1,pet,gle2}. 
Applying the Gibbs conditions to get the equation of state (EOS), 
one may see non-uniform structures of not only 
baryon density but also charge density distribution for the mixed phase, 
and no constant pressure region in EOS. 
If this is the case, the Maxwell construction (MC) 
should become meaningless, where the phase equilibrium between 
two bulk matters with local charge neutrality is assumed. 
The appearance of such SMP in the hadron-quark deconfinement
transition has been expected inside neutron stars and its
implications have been discussed \cite{gle2}. 

The importance of the {\it finite-size effects} in the mixed phase,
which has not been fully included in the previous calculations,
has been emphasized in recent papers: 
especially, the charge rearrangement effect induced 
by the Coulomb interaction should be carefully taken into account.
Actually it has been demonstrated \cite{vos,end1,end2} that 
the Debye screening effect should greatly modify the 
description of the SMP. 

To study non-uniform structure,
we solve the coupled equations of motion based on the density functional
theory \cite{end3},
using the Wigner-Seitz approximation.
In our calculation we can derive the density profiles 
of all the particle species and determine the
configuration of the Coulomb potential {\it exactly} without 
recourse to any approximations included in the recent study. 

\section{Numerical results}

\begin{figure}[htb]
\begin{minipage}[t]{75mm}
\includegraphics[width=75mm]{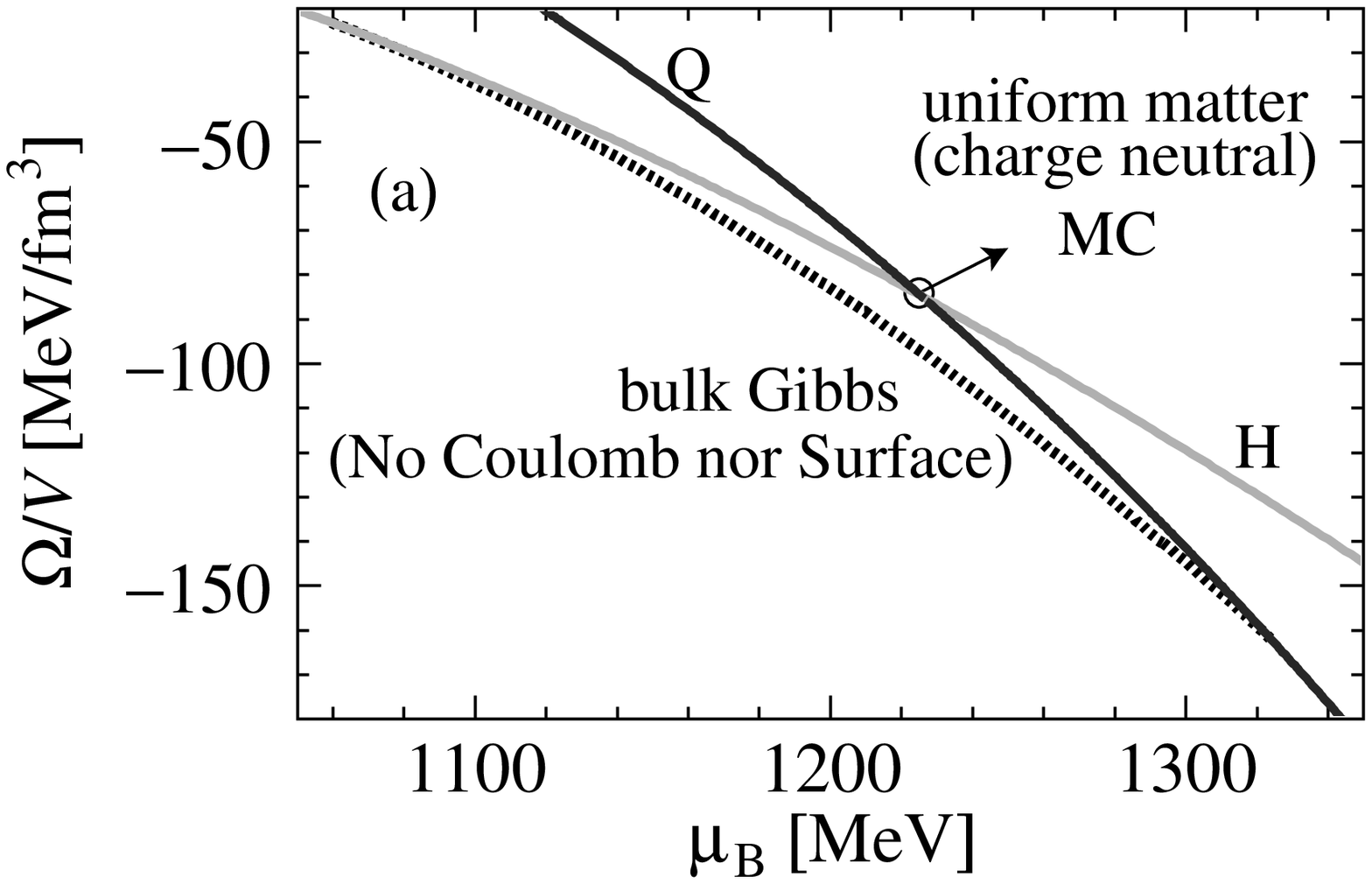}
\caption{Thermodynamic potential density.}
\label{mubome}
\end{minipage}
\begin{minipage}[t]{75mm}
\includegraphics[width=75mm]{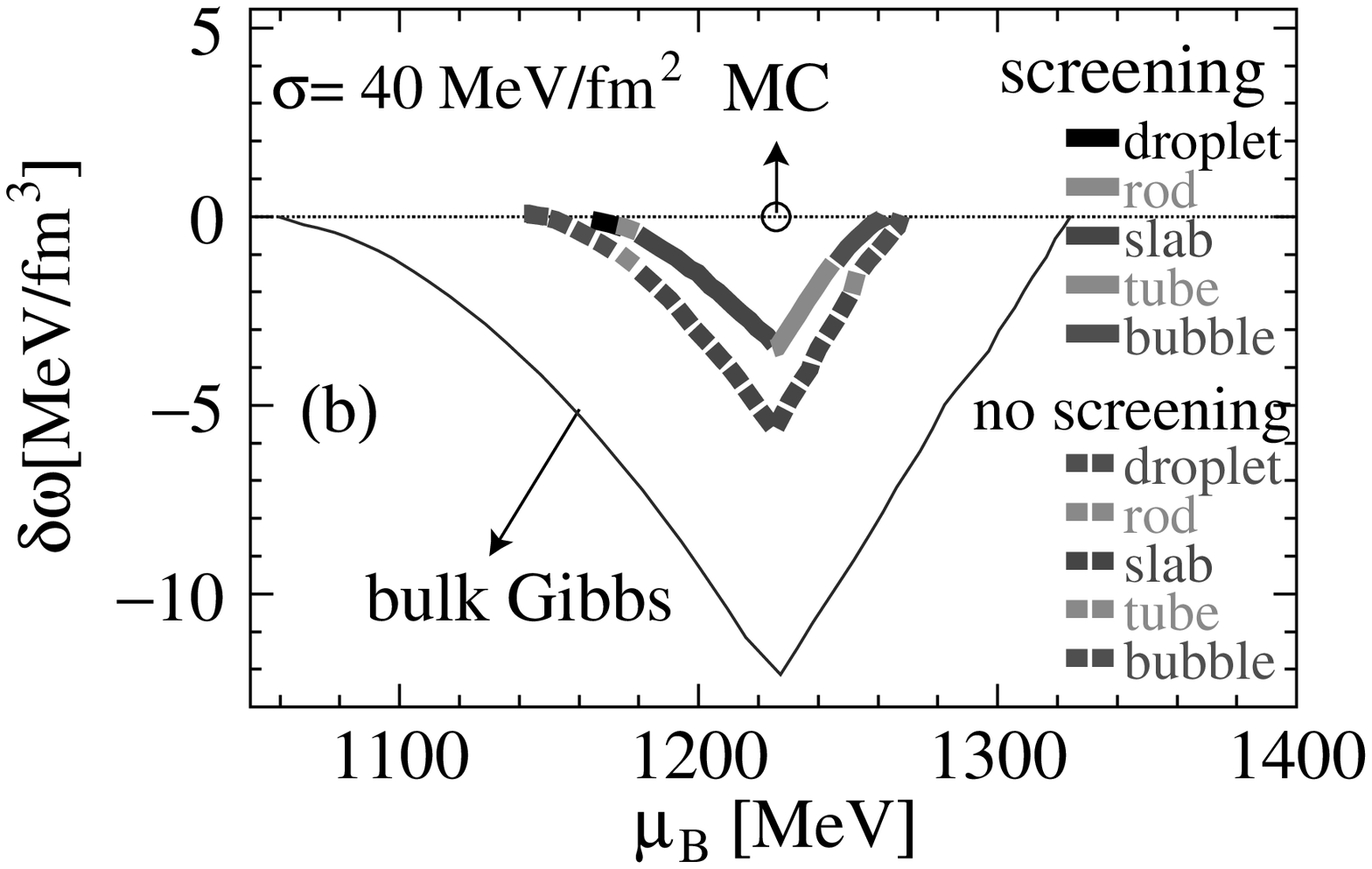}
\caption{Thermodynamic potential density. (a) shows the results
 of each uniform matter and ``bulk Gibbs''. (b) shows the
 difference between the mixed phases and the uniform matter.}
\label{alfo40}
\end{minipage}
\end{figure}

We present the thermodynamic potential density $\omega\equiv\Omega/V$
of each uniform matter 
 in Fig.\ \ref{alfo40} (a) as a function of the baryon number chemical
 potential $\mu_B$: the hadron phase (H) is thermodynamically favorable for 
$\mu_\mathrm{B}$ below 1225 MeV and the quark phase (Q) above it. We
also depict the one denoted by ``bulk Gibbs'' for comparison, 
where the Gibbs conditions are applied but the finite-size effects are
 completely discarded. MC can be represented as a point in this figure. We can
 immediately see
 that ``bulk Gibbs'' smoothly connects H and Q, and the mixed phase
 appears in this wide interval.
We plot $\delta \omega$, the difference of the thermodynamic potential density
between the mixed phase and each uniform matter, in Fig.\ \ref{alfo40}
(b): the curve denoted by ``screening'' is the result of self-consistent
calculation, while the one denoted by ``bulk Gibbs'' corresponds to that
in Fig.\ \ref{alfo40} (a). We also depict another curve denoted by ``no
screening'' to elucidate the charge screening effect, 
which is given by a perturbative
treatment \cite{pet,gle2} of the Coulomb interaction.
Then we can see that the large reduction of the thermodynamical potential from ``bulk Gibbs''
is mainly
given the effect of the surface tension, while the screening effect
further reduces it. 
%
$\delta \omega$ given by MC appears as a point denoted by a circle in
 Fig.\ \ref{alfo40} (b) where two conditions, $P^\mathrm{Q}=P^\mathrm{H}$ and
$\mu_\mathrm{B}^\mathrm{Q}=\mu_\mathrm{B}^\mathrm{H}$, are satisfied.
On the other hand the mixed phase derived from ``bulk Gibbs'' appears in a
wide region of $\mu_{\mathrm{B}}$. 
Therefore, if the region of the mixed phase becomes
narrower, it signals that the properties of the mixed phase become close
to those given by MC. One may clearly see that the thermodynamic potential becomes close to that
given by MC due to the finite-size effects.

\begin{figure}[htb]
\begin{tabular}{c c}
\begin{minipage}[t]{75mm}
\includegraphics[width=75mm]{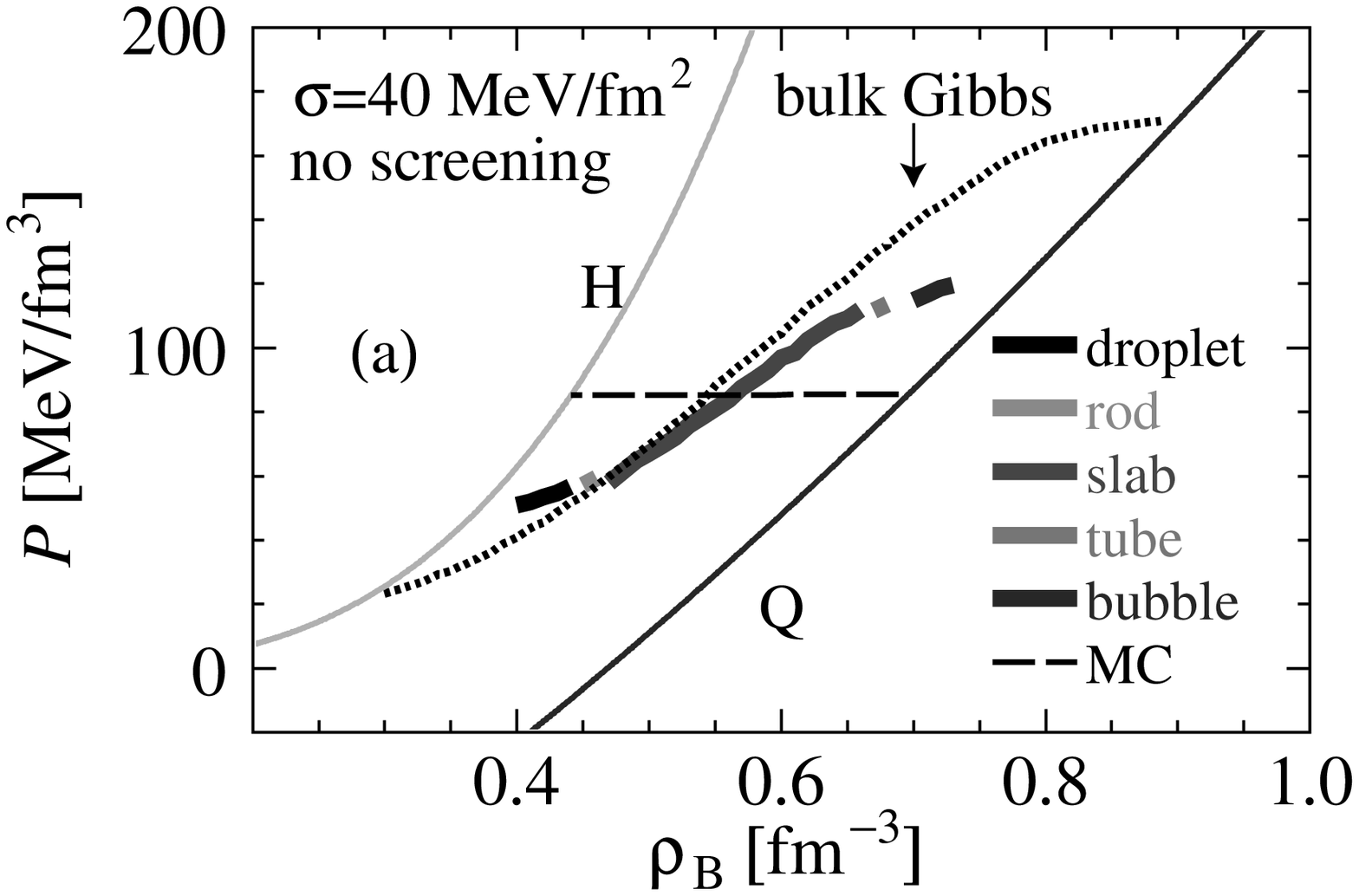}
\caption{Without screening (No Coulomb).}
\label{pres40no}
\end{minipage}
\begin{minipage}[t]{75mm}
\includegraphics[width=75mm]{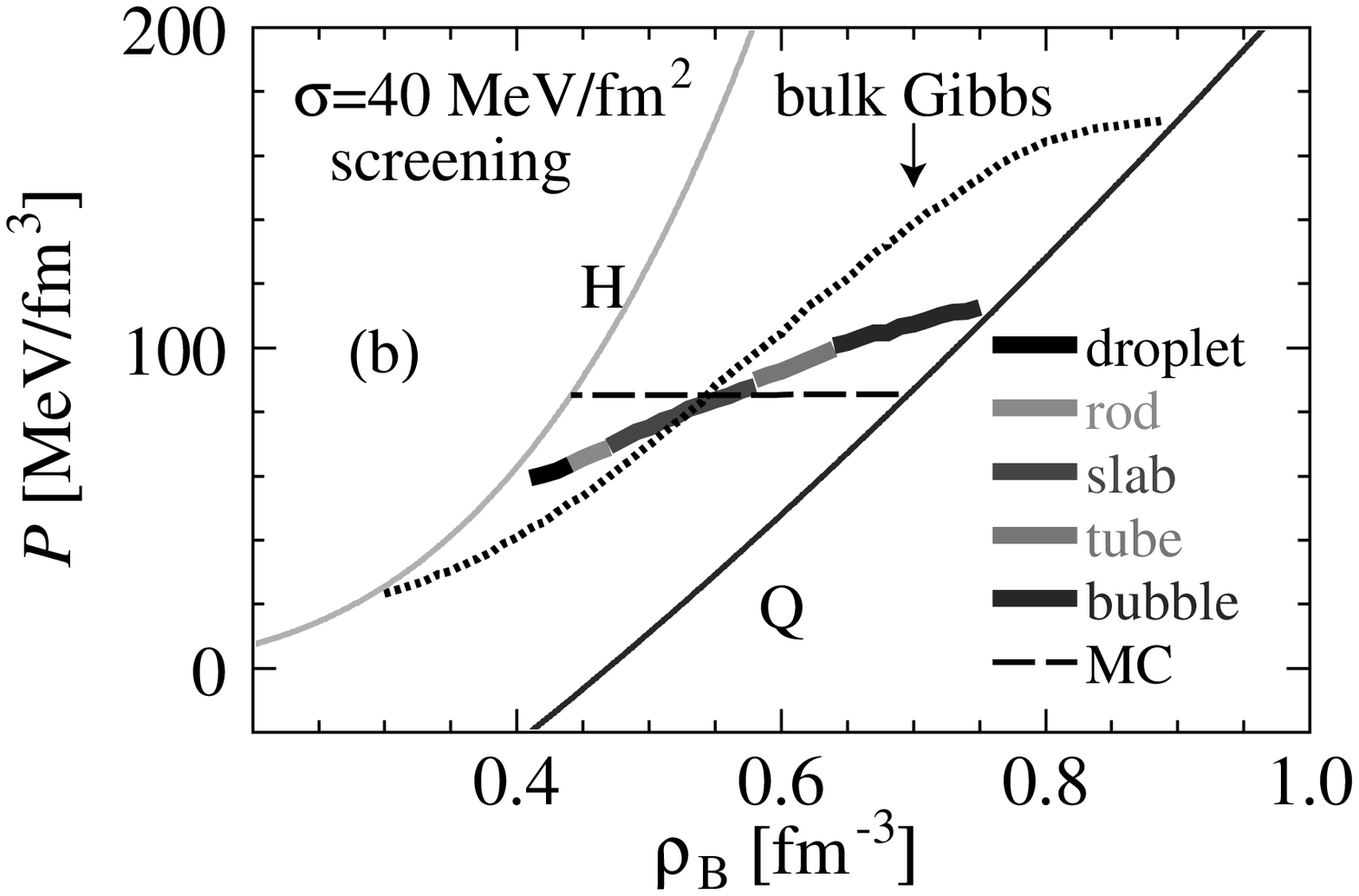}
\caption{Pressure as a function of baryon-number density. (a) is the
 result of ``no screening'' and (b) ``screening''. The results given
 by ``bulk Gibbs'' and MC are also presented for comparison. }
\label{pres40}
\end{minipage}
\end{tabular}
\end{figure}

Figures.\ \ref{pres40} (a) and (b) show EOS in the cases 
with and without the screening effect. 
The pressure becomes more similar to that
given by MC by the finite-size effects. 
Moreover,  in Fig.\ \ref{pres40} (b), it becomes more close to
MC by the charge screening effect, 
which shows a larger pressure in the beginning 
and weaker one in the end of the phase transition.

\newpage

\section{Summary and Concluding remarks}

We have seen that the finite-size effects changes the properties of
the hadron-quark mixed phase which is expected in hybrid stars.
In particular, the region in the baryon-number chemical potential 
is restricted by the charge screening effect. 
We have seen that EOS becomes close to that
with MC by the finite-size effects; 
EOS becomes more similar to that
with MC by the charge screening effect.

Let us briefly consider some implication of these our results for compact star phenomena.
Glendenning and Pei \cite{gle2} suggested many SMPs appear
in the core region by using ``bulk Gibbs'':
the mixed phase should appear for several kilometers. 
However we can say that the region of SMP
should be narrow in the $\mu_\mathrm{B}$ space and EOS is more similar
to that of MC 
due to the finite-size effects.
These results seem to be consistent with those given by other studies.
Bejger et al.\ \cite{bejg} have
examined the relation between the mixed phase
and glitch phenomena, and shown that the mixed phase should be narrow if the
glitch is generated by the mixed phase in the inner core. 
On the other hand the gravitational
wave asks for density discontinuity in the core region \cite{mini}. 
It is very interesting to study the relation between these phenomena and
our results in more detail.







\end{document}